\documentclass[reprint,superscriptaddress,showpacs]{revtex4-1}
\usepackage{graphicx}

\usepackage[latin1]{luainputenc}

\usepackage{hyperref}

\usepackage{SIunits}
\usepackage{bm}

\renewcommand{\vec}[1]{\ensuremath{\boldsymbol{#1}}}

\begin{document}

\title{Is magnetic chiral dichroism feasible with electron vortices?}
\author{P. Schattschneider}
\affiliation{Institut f\"ur Festk\"orperphysik, Technische Universit\"at Wien, A-1040 Wien, Austria}
\author{S. L\"offler}
\author{M. St\"oger-Pollach}
\affiliation{USTEM, Technische Universit\"at Wien, A-1040 Wien, Austria}
\author{J. Verbeeck}
\affiliation{EMAT, University of Antwerp, Groenenborgerlaan 171, B-2020 Antwerp, Belgium}

\begin{abstract}
We discuss the feasibility of detecting magnetic transitions with focused electron vortex probes, suggested by selection rules for the magnetic quantum number. We theoretically estimate the dichroic signal strength in the L$_{2,3}$ edge of the ferromagnetic d metals. It is shown that under realistic conditions, the dichroic signal is undetectable for nanoparticles larger than $\sim \unit{1}{\nano\meter}$. This is confirmed by a key experiment with nanometer sized vortices. 
\end{abstract}

\maketitle

After the publication of two seminal papers~\cite{UchidaNature2010,VerbeeckNature2010}, electron vortex beams have attracted considerable interest~\cite{McMorranScience2011, Bliokh2012, Schattschneider2012,LoefflerACTA2012,Verbeeck20131114,Guzzinati2013,Lubk2013}. Vortex beams are free electrons carrying orbital angular momentum (OAM). Their potential ranges from probing chiral specimens with elastic or inelastic scattering over the manipulation of nanoparticles~\cite{Verbeeck20131114}, clusters and molecules to the study of magnetic properties. Experimental evidence of the detection of chirality in electronic transitions~\cite{VerbeeckNature2010} led to the suggestion~\cite{LloydPRL2012} that electrons with topological charge are better probes for such experiments than the plane waves in the scattering geometry for detecting energy loss magnetic chiral dichroism (EMCD)~\cite{SchattNature2006}. 

However, care must be taken when comparing vortices to plane wave electron probes. Results depend sensitively on the experimental parameters such as convergence and collection angles, position of the holographic mask, etc. Here, we discuss the dichroic L$_{2,3}$ dipole transitions in the 3d ferromagnets --- a standard for EMCD experiments in the electron microscope~\cite{SchattschneiderJAP2008} --- mediated by an incident electron with topological charge. 
 
The most dominant contributions to the electron energy loss spectrometry (EELS) signal are electric dipole transitions. Higher multipole transitions have low transition amplitudes contributing less than \unit{10}{\%} at the scattering angles of $< \unit{20}{\milli\rad}$ relevant in EELS~\cite{Manson1972,U_v111_i_p1163,PRB_v40_i4_p2024}. 

In case of an L edge dipole transition which changes the magnetic quantum number of an atom located at the vortex center by $\mu$, an incident electron $\psi_m(\vec{r})=e^{im\varphi} f(r)$ with topological charge $m$ transforms into an outgoing
wave~\cite{SchattschneiderPRB2010}
\begin{equation}
\psi_{m, \mu}(\vec{r})=e^{i(m+\mu)\varphi_r} f_\mu(r) f(r),
\label{Atom}
\end{equation}
where $\varphi_r$ is the azimuthal angle, and 
\begin{equation}
f_\mu(r)=
\frac{i^{\mu}}{2 \pi}q_E^{1-|\mu|} \int_0^\infty \frac{q^{1+|\mu|} J_{|\mu|}(q r) \langle
j_1(Q) \rangle_{ELSj}}{Q^3} dq,
\label{psi2}
\end{equation}
with $\langle
j_1(Q) \rangle_{ELSj}$ the matrix element of the spherical Bessel function between initial and final radial atomic wave functions, and $Q^2=q^2+q_E^2$. Here, $q$ is the transverse scattering vector that relates to the experimental scattering angle $\theta$ as $q=k_0 \theta$, and $\hbar q_E$ is the scalar difference of linear momenta of the probe electron before and after inelastic interaction, also known as the characteristic momentum transfer in EELS~\cite{Egerton}.

\begin{figure}[ht]
	\centering
	\includegraphics[width=0.4\textwidth]{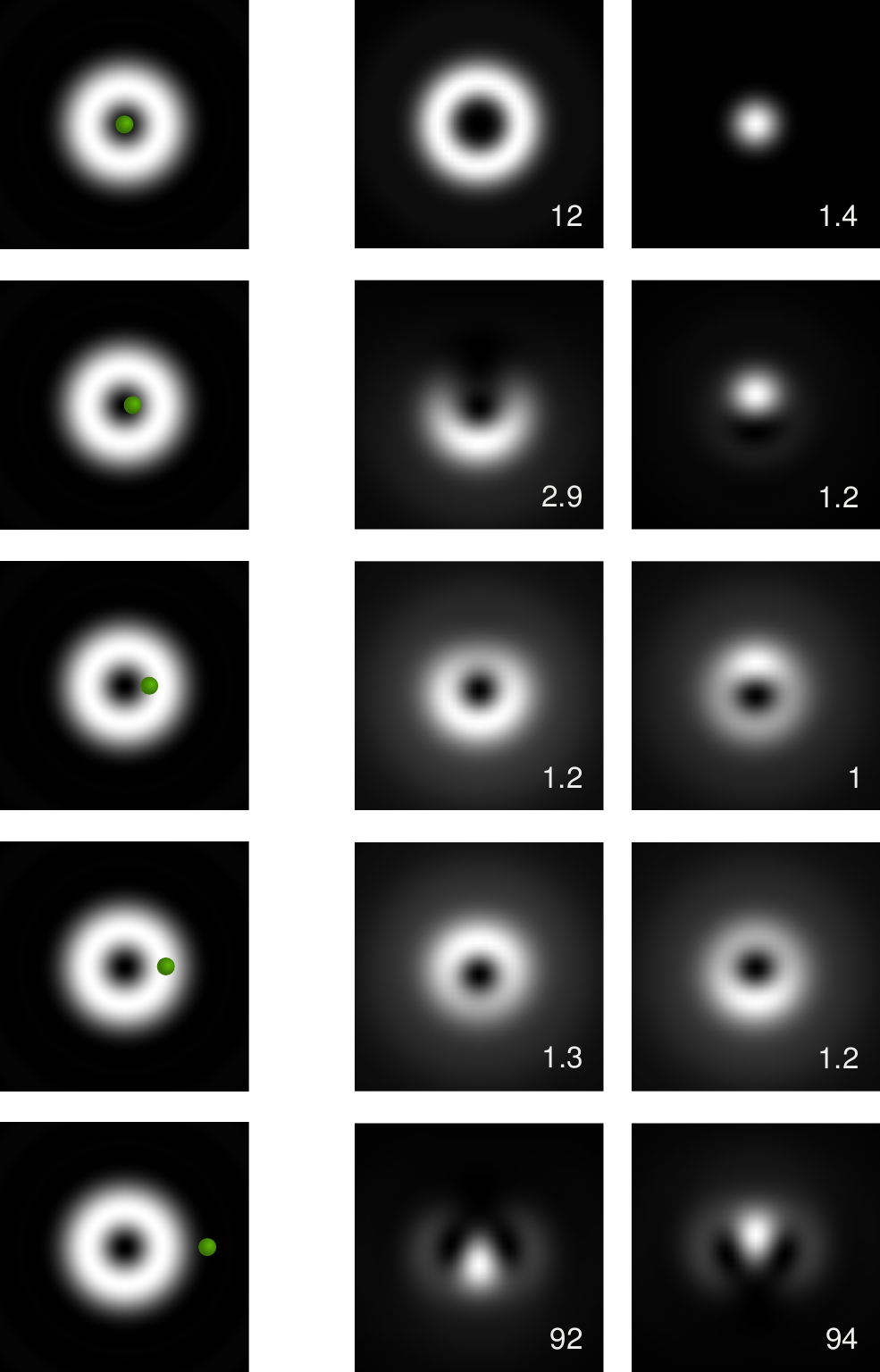}
	\caption{Left column: incident $m=1$ vortex with ring radius of \unit{0.9}{\nano\meter} and displaced atoms (green spheres: 0, 0.2, 0.6, 1, and \unit{2}{\nano\meter} from the vortex center). Middle and right columns: energy filtered diffraction patterns of atomic chiral transitions with $\mu=-1$ for incident waves with $m=1$ (middle) and $m=-1$ (right) in the Fe L edge. The values in the right bottom corners give the respective scaling factors for the intensities. The intensities in the lowermost panels are about 90 times weaker than those in the middle panels. The panels map scattering angles of $\pm \unit{10}{\milli\radian}$.}
	\label{fig:DecenterFocDP}
\end{figure}

When there are several transition channels at the same energy, the outgoing probe electron is in a mixed state, described by a reduced density matrix. The total intensity is a sum over intensities in the respective channels
\begin{equation}
 I_m(\vec{r})=\sum_\mu|\psi_{m,\mu}(\vec{r})|^2.
\label{Int}
\end{equation}
The dichroic signal is measured in the diffraction plane. It can readily be calculated via Fourier transforming Eqs.~\ref{Int} and \ref{Atom}. According to a well-known theorem for the Fourier-Bessel transform of a function of azimuthal variation $e^{im \varphi}$, 
one has
\begin{equation}
\tilde \psi_{m,\mu}(\vec{q})= 
\frac{i^{m+\mu}}{2 \pi} e^{i(m+\mu)\varphi_q}\int_0^{\infty} f_\mu(r) f(r)J_{m+\mu}(q r)\,  r dr.
\label{FT2}
\end{equation}
The outgoing electron has topological charge $m+\mu$. The radial intensity profiles $|\tilde \psi_{m,\mu}(\vec{q})|^2$ of the inelastically scattered vortex with $m=1$ in the diffraction plane for transition channels $\mu=\pm 1$ in the Fe L$_3$ edge are shown in Fig.~\ref{fig:DecenterFocDP}. The figure shows how chiral transitions in ferromagnetic specimens can be selected with a collection aperture subtending the innermost region. Note that this region corresponds to the characteristic momentum transfer for Fe L, 0.24~at.u. (equivalent to a scattering angle of $\sim \unit{2.5}{\milli\radian}$ at \unit{200}{\kilo\electronvolt} incident electron energy). For an atom centered in the vortex, these profiles closely resemble those of helical waves with winding number $m+\mu$. This is the basis for probing magnetic transitions with vortex electrons.

When the excited atom is at a distance $R$ from the vortex center, the incoming wave must be expanded into cylindrical eigenfunctions over the atom position. In \cite{Comment}, an incident Bessel beam was assumed and expanded according to the addition theorem of Bessel functions~\cite{Abramowitz1965}.

However, in the experiment, the vortex impinging on the atom is not a Bessel beam but rather an aperture limited convergent spherical wave (here corresponding to a convergence semi-angle $\alpha=\unit{1.2}{\milli\radian}$) with topological charge $m= \pm1$. In this case, it is more convenient to expand the wave function into cylindrical harmonics around the atom center~\footnote{Any function can be expanded into Bessel functions~\cite{SchattUM2011} and then the addition theorem can be applied.}. Upon angular momentum expansion~\cite{FrankeArnold2004}, we obtain a Fourier series in the azimuthal angle $\varphi_r$,
\begin{equation}
\psi_m(\vec{r}-\vec{R})=\sum_l a^{m}_l(r) e^{il \varphi_r},
\label{psiR}
\end{equation}
where the coefficients are functions of the convergence semi-angle $\alpha$ and the atomic displacement $\vec{R}$, $a^{m}_l(r)=a^{m}_l(\alpha,\vec{R},r)$. We note in passing that the largest coefficient will be that for $l=m+\mu$, and that in the limit ${R \to 0}$, all other coefficients vanish. Eq.~\ref{psiR} shows clearly that the outgoing electron wave is a coherent superposition of angular momentum eigenstates. This is a consequence of the uncertainty relationship for OAM and angular position: the interaction restricts the outgoing inelastically scattered electron to the extension $r_z$ of the atomic orbitals implied in the electronic transition. Seen from the vortex center, this translates into an uncertainty of the azimuthal angle $\delta \varphi \approx 2 r_z/R$, and $\delta L_z \geq \delta \varphi^{-1}$.
It follows that $ L_z $ is not a constant of motion any more. This is an important difference to optical absorption spectroscopy where the selection rules are governed by the transfer of spin angular momentum (SAM) which is position independent. In EELS, however, they are governed by the transfer of OAM which is position dependent~\footnote{This can be understood from the parallel axis theorem; the angular momentum depends on the reference frame. The quantity that is independent of the reference frame is the vorticity~\cite{Lubk2013} which cannot be measured with the present experimental setup.}.

Application of Eq.~\ref{FT2} to the Fourier coefficients $a_l^{m,\mu}$ results in the diffraction plane representation
\begin{equation}
\tilde \psi_{m,\mu}(\vec{q})=
\sum_l\frac{i^{l}}{2 \pi} e^{il\varphi_q}
\tilde a^{m,\mu}_l(q)
\label{FT3}
\end{equation}
with
\begin{equation}
\tilde a^{m,\mu}_l(q)=\int_0^{\rho} a^{m, \mu}_l(r)f_\mu(r)J_{l+\mu}(q r) r dr .
\label{cj}
\end{equation}
Numerically, the upper integration limit is determined by the extension $\rho$ of the atomic function $f_\mu(r)$. For the following calculations, we assumed a rather large interaction radius $\rho=10$~at.u. where $f_\mu$ is sufficiently small to be used as a cutoff for the Fourier transform. Results for different displacements of the atom from the vortex center (ring radius \unit{0.9}{\nano\meter}, corresponding to a convergence semi-angle of \unit{1.2}{\milli\radian} at \unit{200}{\kilo\electronvolt} beam energy) are shown in Fig.~\ref{fig:DecenterFocDP}. 

It is evident that the symmetry breaking responsible for the EMCD effect survives only up to displacements below \unit{1}{\nano\meter}. Beyond that value, the diffraction patterns for left- and right handed chiral vortices (middle and right columns) are practically indistinguishable. It must be noted that the atoms close to the vortex center (which show the highest difference) contribute the faintest signals because $\lim_{r\to 0} f(r)=0$. 

The EMCD signal is defined as the relative difference of signals from vortices with $m=\pm 1$
\begin{equation}
	\text{EMCD}=2 \cdot \frac{I_{+1}-I_{-1}}{I_{+1}+I_{-1}}.
	\label{EMCD}
\end{equation}
The independent variables have been omitted for clarity.
For fully spin-polarized systems, one has
\[
I_m=\sum_{\mu=-1}^1 C_m^\mu |\psi_{m \mu}|^2
\]
where $C_m^\mu $ are derived from the Clebsch-Gordan coefficients~\cite{SchattPRB2012,mdff-factorization}. 

\begin{figure}[ht]
	\centering
	\includegraphics[width=0.4 \textwidth]{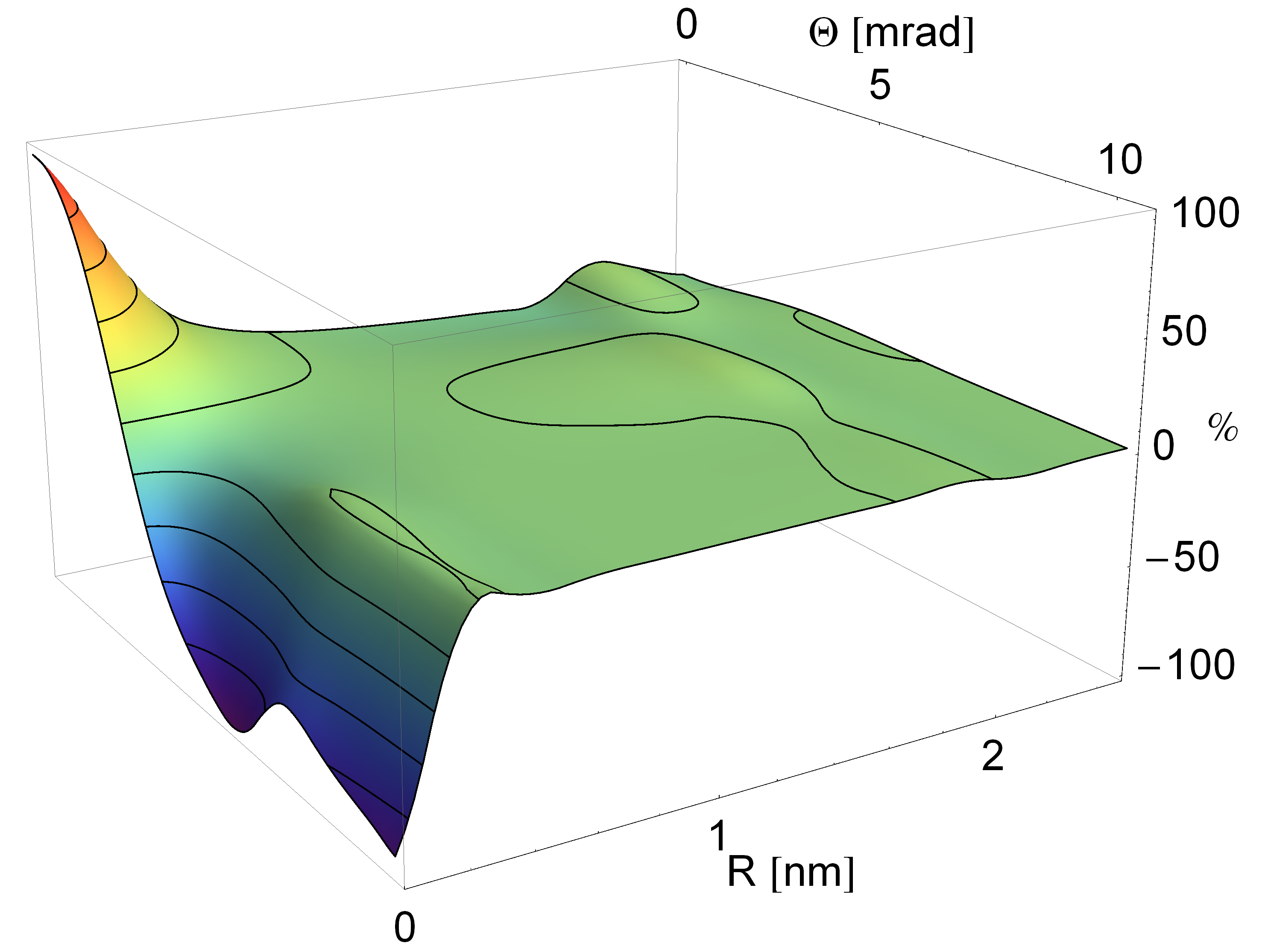} \\
	\includegraphics[width=0.4 \textwidth]{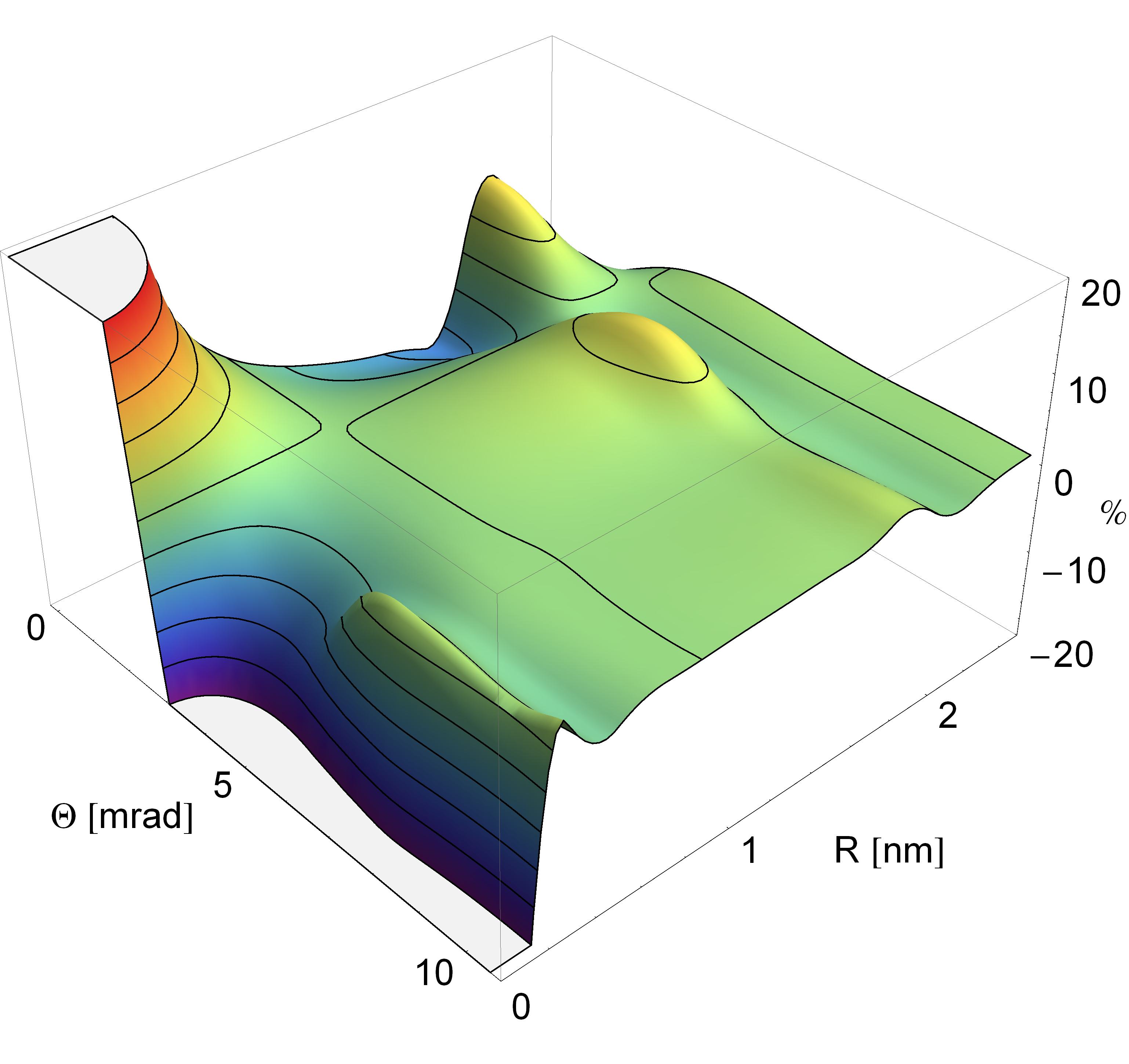}
	\caption{EMCD signal for an Fe~L edge assuming a collection semi-angle of \unit{1.2}{\milli\radian} as a function of scattering angles and atom displacements (up to \unit{2.5}{\nano\meter}). The isolines trace an increase of \unit{20}{\%} each. The lower panel is a zoom with isolines every \unit{4}{\%} of EMCD signal.}
	\label{fig:AvgFocDP}
\end{figure}

When a homogeneous specimen is illuminated, all atoms will contribute incoherently with their respective signals. The expected energy filtered diffraction pattern will then be radially symmetric. It is obtained as the integral of the radial $\theta$-traces over all azimuths and all displacements $R$. Defining the collection semi-angle $\beta$ of the detector, the signal from a vortex with charge $m$ is 
\begin{equation}
I_m(\beta)=\int_0^\beta \int_0^{R_{max}} \bar I_m(R,\theta) \, d^2R \, \theta \, d\theta 
\label{Iq}
\end{equation}
with
\begin{equation}
\bar{I}_m(R,\theta)=\frac{1}{2 \pi}\int_0^{2 \pi}I(\vec{R},\theta,\varphi_\theta) \, d\varphi_\theta.
\label{barI}
\end{equation}
$\bar{I}_m$ is shown in Fig.~\ref{fig:AvgFocDP} for varying displacements and scattering angles. It is the average contribution to the EMCD signal of an atom displaced from the vortex center by $R$, independent of its azimuth.
 
\begin{figure}[ht]
	\centering
	\includegraphics[width=0.4 \textwidth]{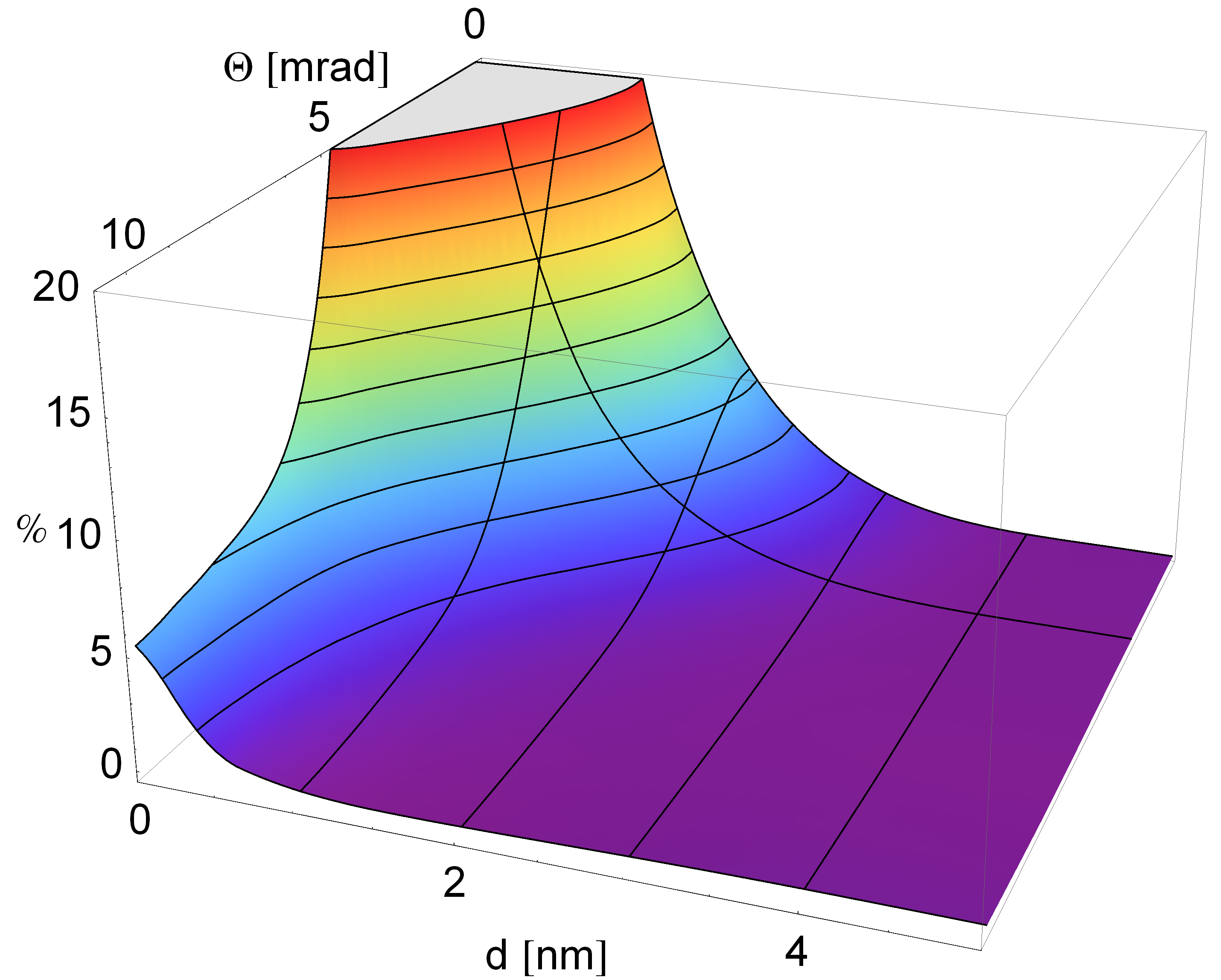}
	\caption{Integrated EMCD signal for an Fe~L edge assuming a convergence semi-angle of \unit{1.2}{\milli\radian} for disk-like nanoparticles (up to \unit{5}{\nano\meter} diameter) which are centered in the vortex, as a function of collection angle. The isolines trace increments of \unit{2}{\%}. At a diameter of \unit{3}{\nano\meter}, the EMCD signal has dropped below \unit{2}{\%} for zero collection angle.}
	\label{fig:RelDiff}
\end{figure}

Fig.~\ref{fig:AvgFocDP} is consistent with Fig.~\ref{fig:DecenterFocDP}: for displacements larger than \unit{0.5}{\nano\meter}, the diffraction patterns start to be indistinguishable, and this is also where the EMCD signal drops below noise level. More precisely, as shown in the lower panel, it drops below \unit{4}{\%} for displacements as small as \unit{0.6}{\nano\meter}, even for the smallest scattering angles. Interestingly, for scattering angles larger than about \unit{3}{\milli\radian}, the EMCD signal changes sign. This can be understood from the contrast inversion in the angular scattering profiles of the centered atom in Fig.~\ref{fig:DecenterFocDP}.  Larger collection angles should therefore be avoided, in order to avoid diminishing the signal. The integrated EMCD signal of a nanoparticle of diameter $d$ as a function of collection angle, obtained from Eq.~\ref{Iq}, is shown in Fig.~\ref{fig:RelDiff}.
 
Even for zero collection angle --- where the EMCD effect is strongest ---, the signal drops below \unit{2}{\%} for particles larger than $\sim$~\unit{2.5}{\nano\meter}. The best signal-to-noise ratio was calculated to be at collection semi-angles of about \unit{3}{\milli\radian}, again to be understood from Fig.~\ref{fig:DecenterFocDP}: at this $\beta$, the difference signal for the centered atom is largest. With this setup, an EMCD signal of $>\unit{5}{\%}$ (which is a realistic threshold for detection) can only be detected for particles smaller than \unit{1.5}{\nano\meter}. 
 
\begin{figure}[ht]
	\centering
	\includegraphics[width=0.45\textwidth]{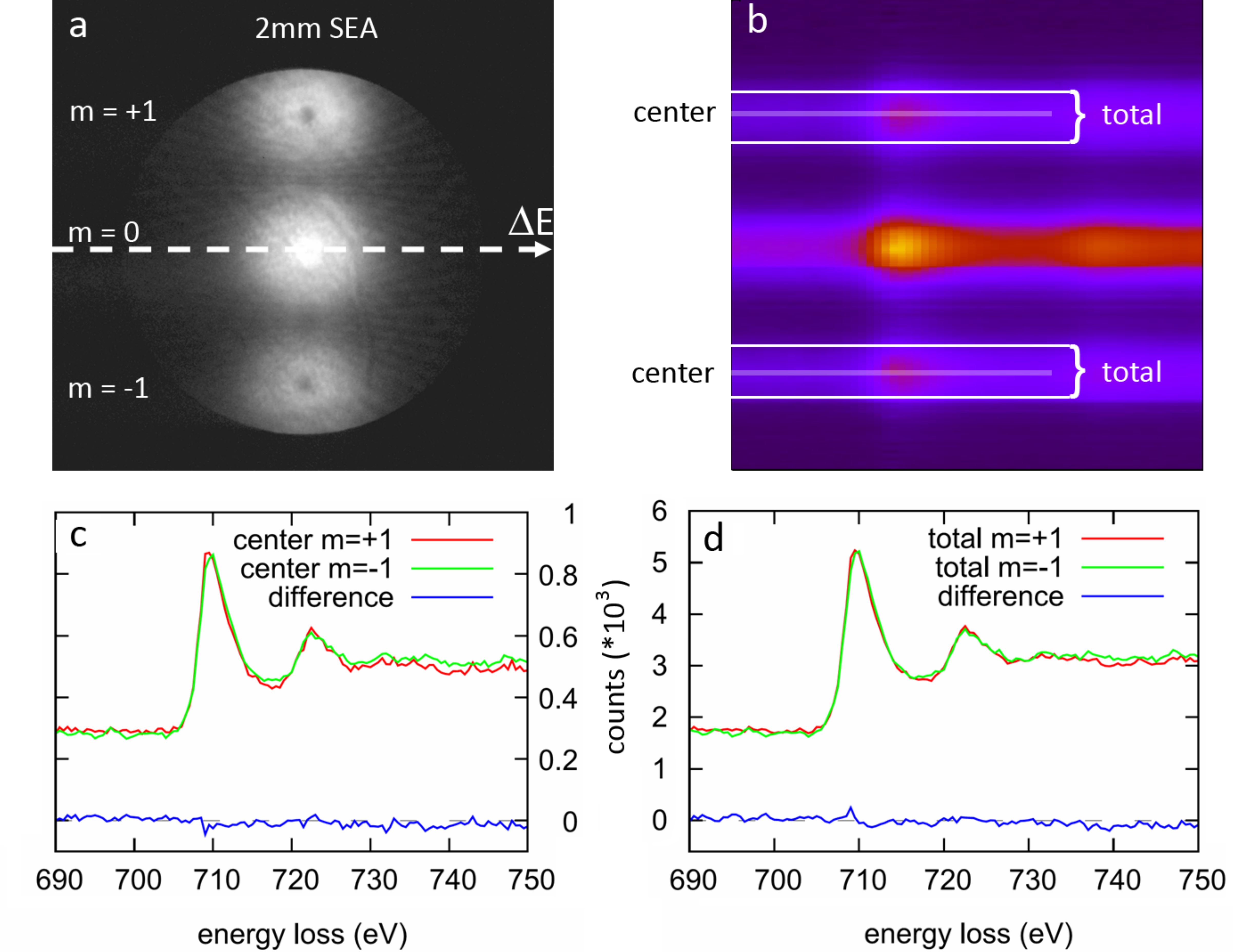}
	\caption{Experiment investigating the feasibility of EMCD detection: a) electron vortices after passing the sample and the SEA. b) raw spectrum image of the Fe-L$_{2,3}$ edge. The centers of the vortices and the total signals are labeled. c) raw spectra and the respective differences of the vortex centers. d) raw spectra summing over the total vortex intensities and the respective differences.}
	\label{fig:resultVienna}
\end{figure}

Several experiments were performed with a variety of vortex diameters and materials, but none of them showed an EMCD signal.  Shown here as an example is an experiment using an electro-chemically etched iron specimen of \unit{80}{\nano\meter} thickness. The vortices were created using the a convergence semi-angle of \unit{1.2}{\milli\radian}. The collection semi-angle was chosen to be \unit{2.8}{\milli\radian}. The electron vortices passed through the specimen into the \unit{2}{\milli\meter} spectrometer entrance aperture (SEA) --- Fig.~\ref{fig:resultVienna}a --- and were subsequently deflected by the magnetic prism of a GATAN GIF Tridiem attached to a FEI TECNAI F20 forming a spectrum image --- Fig.~\ref{fig:resultVienna}b. Finally, the raw spectra (without any background subtraction or intensity normalization) was compared. No EMCD effect was detected, as shown in Figs.~\ref{fig:resultVienna}c and ~\ref{fig:resultVienna}d.

In conclusion, we find theoretically and experimentally that  EMCD with incident focused vortex electrons~\cite{LloydPRL2012} is ineffective for particles lager than a couple of nanometer. The signal drops rapidly below \unit{2}{\%} even for the smallest collection angles. With present instruments, it is therefore virtually impossible to detect chiral dichroism in the discussed scattering geometry. The situation is probably more favorable for atom-sized vortices which have the additional advantage of channeling along the atomic columns~\cite{LoefflerACTA2012, Lubk2013}, but this discussion is beyond the scope of the present paper.

Experimental evidence of EMCD spectroscopy with electron vortex beams reported previously~\cite{VerbeeckNature2010} was based on a different geometry, namely an incident converging wave and a strong defocus of the holographic mask sitting below the objective lens. This mask acted as a discriminator for the topological charge of the outgoing electrons. It should be noted that these observations pose severe limits to medium scale EMCD in the discussed scattering geometry but they do not exclude the possibility of EMCD with vortex probes of atomic scale, or with different geometry.

\begin{acknowledgements}
The authors thank Tomasz Wojcik for kindly providing the iron specimen. P.S. and S.L. acknowledge financial support of the Austrian Science Fund, project I543-N20. J.V. was supported by funding from the European Research Council under the 7th Framework Program (FP7), ERC grant No.~246791 COUNTATOMS, ERC Starting Grant No.~278510 VORTEX and a contract for an Integrated Infrastructure Initiative, reference No.~312483-ESTEEM2.
\end{acknowledgements}

\bibliography{Lit2012}

\end{document}